\documentclass[secnumarabic, graphics,floatfix,prl]{revtex4-1}
\usepackage{amsmath}
\usepackage{amssymb}
\usepackage{graphicx}
\usepackage{color}
\usepackage{float}
\usepackage[ansinew]{inputenc}
\usepackage[dvipsnames]{xcolor}

\def\blk{\color{black}}

\begin{document}

\title{Magneto-Optical Functions at the 3$p$ resonances of Fe, Co, and Ni: \emph{Ab-initio} description and experiment}

\author{F. Willems$^1$}
\email{willems@mbi-berlin.de}
\author{S. Sharma$^1$}
\email{sharma@mbi-berlin.de}
\author{C. v. Korff Schmising$^1$}
\author{J. K. Dewhurst$^2$}
\author{L. Salemi$^3$}
\author{D. Schick$^1$}
\author{P. Hessing$^1$}
\author{C. Str\"uber$^1$}
\author{W. D. Engel$^1$}
\author{S. Eisebitt$^{1,4}$}
\affiliation{$^1$ Max Born Institute for Nonlinear Optics and Short Pulse Spectroscopy, Max-Born-Strasse 2A, 12489 Berlin, Germany}
\affiliation{$^2$ Max-Planck-Institute for Microstructure Physics, Wienberg 2, 06120 Halle(Saale), Germany.}
\affiliation{$^3$ Department of Physics and Astronomy, Materials Theory, Uppsala University, 75120 Uppsala, Sweden}
\affiliation{$^4$ Institut für Optik und Atomare Physik, Technische Universität Berlin, 10623 Berlin, Germany}

\date{\today}
\begin{abstract}
We present theoretical and experimental data on the magneto-optical
contributions to the complex refractive index in the extreme ultraviolet (XUV) range covering the 3$p$ resonances of Fe, Co, and Ni. Comparing the spectra from density functional theory  with magnetic circular dichroism  measurements we find that many body corrections and
local field effects are of crucial importance for accurate description of the spectra. Our results are relevant for the application of static XUV spectroscopy for multi element magnetic systems as  well as for the investigation of ultrafast magnetization dynamics. 
\end{abstract}



\maketitle

X-ray spectroscopy is a very powerful tool as it can probe electronic and magnetic structure for each constituent element of a material separately via transitions from core levels to unoccupied states\cite{Stohr2006}. Magnetic order is probed via dichroism which is a contrast based on the projection vector of the X-ray polarization onto the sample magnetization and the element selectivity stems from the increased scattering cross section when the incident X-ray photon energy is tuned to resonate with transitions from core levels, which are energetically separated for different atoms and in different chemical environment.  For the important class of magnetically ordered materials containing 3$d$ transition metals (TMs), the use of $L$-edge (2$p$ core levels) spectroscopy at synchrotron facilities was dominant to determine magnetic properties, due to the large magnitude of the magnetic dichroism at these transitions and the ability to apply sum rules to disentangle spin and orbital contributions to the magnetic moments\cite{Stohr2006,Chen1995}. In recent years, however, the use of magnetic dichroism in the extreme ultraviolet regime, exploiting the $M$-edge resonances (3$p$ core levels) in 3$d$ TMs, has strongly increased due to the growing availability of laboratory based high harmonic generation (HHG) sources\cite{Vodungbo2011,Kfir2015,Kfir2016,Hickstein2015,Schmising2017,Ellis2018}. 

The major drawback of these   $M$-edge experiments is the challenging data interpretation, because (a) the strong overlap of the spin-orbit split core-levels ($3p_{\frac{3}{2},\frac{1}{2}}$) implies that the optical sum-rules cannot be applied  to disentangle  spin and orbital contributions in the total angular moment  and (b) the $M$-edge resonances of  the 3$d$ TMs exhibit  substantial overlap with each other for photon energies above and in particular well below the absorption edges\cite{Kfir2015,Hickstein2015,Mathias2012,Turgut2013,Radu2015}. This has led to several unresolved controversies in the interpretation of the experimental data  to reconcile which  one requires a  fully parameter free theoretical description to base the interpretation of the experimental data on. However, all previous calculations of $M$-edge spectra even for the for the most common 3$d$ TMs (Fe, Co, and Ni) rely on \emph{ad-hoc} Gaussian broadening, energy shifts and amplitude scaling\cite{Valencia2010, Turgut2016,Zusin2018} to bring theoretical results close to experiments.

In the present work we combine experimental and theoretical studies for the 3$d$ TMs Fe, Co, and Ni at their $M$-edge resonances. Experimentally, we measured both the dispersive and absorptive parts of the magneto-optical (MO) contributions to the refractive index, eliminating the need to complement information by a Kramers-Kronig analysis\cite{Valencia2006}, which is inherently inaccurate when applied to rapidly changing functions measured in a small energy widow. Theoretically, we provide a fully \emph{ab-initio} description of these spectra. 
The important breakthrough of this work is that we find the \emph{ab-initio} calculations to be in unprecedentedly good agreement with the accurate experimental results. With this we are also able to provide the fundamental reasons behind previous discrepancies between theory and experiment. Most importantly, we are able to assign physical processes that lead to each feature in the experimental spectra, a prerequisite to disentangle signals from overlapping XUV-MO-based spectra in multi-component materials. This will have direct and significant consequences not just for static but also dynamic spectral studies, e.g. performed in the fields of femtomagnetism and spintronics.


The experiments were carried out at the BESSY II synchrotron facility on the beamline UE112-PGM-1, where we complemented the ALICE\cite{Abrudan2015} end-station by a home-built polarization analyzer. The variable polarization of incident XUV radiation enabled two types of measurements in the same setup, both on the same XUV-transmissive thin film samples: XUV-magnetic circular dichroism (MCD) measurements as well as detection of the Faraday rotation, from which we retrieved the absorptive and the dispersive part of the MO-functions, respectively. The monochromator energy resolution was set to E/$\Delta$E = 30000 while preserving a  degree of polarization given by the Stokes parameters of $S_3 = 0.99$, and $S_2 = 0.99$\cite{Bahrdt2010} at the sample for circular and linear polarization, respectively. 
Magnetron sputtering was used  to grow 15\ nm thick layers of Fe, Co, and Ni on $ {Si_{3}N_{4}}$ membranes of 20\ nm thickness, capped by a 3\ nm Al layer oxidation protection. All of the films have their magnetic easy axis in the sample plane and exhibit a coercivity of $< 10$\ mT.

In the XUV-MCD measurement we recorded the transmitted intensity ($ {I_\pm}$) of circularly polarized light through the sample for two magnetic field directions (+,-) as a function of photon energy ($\hbar\omega $). We calculate the XUV-MCD as $ D= 0.25 \log(I_+/I_-)$.
In the Faraday rotation measurement linearly polarized XUV light is transmitted through the sample. The reflected intensity from our analyzer $ {I_{R\pm}} $ is recorded for two magnetization directions (+,-), while the analyzer angle is set to $\alpha=45{^\circ}$ with respect to the polarization plane. The Faraday rotation angle $\Phi_{F}$ as function of incident photon energy is retrieved via the magnetic asymmetry $ {A=(I_{R+}-I_{R-})/(I_{R+}+I_{R-})}$ as  $\Phi_{F} = 1/2 \arcsin (A/P)$, where $P$ is the polarizing power of the polarimeter.
In both measurements, the glancing angle of incidence on the sample was $50^\circ$ and the sample magnetization was set via an in-plane electromagnet. 
By correcting for the angle of incidence and non-negligible refraction of the XUV light at the vacuum-sample interface, we retrieve the elemental MO-functions independent of the experimental geometry. 

Following Valencia at el.\cite{Valencia2006}, we write the complex refractive index $n$ for the two circularly polarized Eigenmodes (+,-) as
\begin{equation}
n_{\pm}(\omega)=1-(\delta (\omega) \pm \Delta \delta (\omega)) + i (\beta (\omega) \pm \Delta \beta (\omega)),
\end{equation}
The relation between the MO-functions ($\Delta\delta(\omega), \Delta\beta(\omega)$) and the measured quantities D and $\Phi_{F}$ is in accordance with the work of Kunes et al.\cite{Kunes2001} and written in Eq.~(\ref{eq:MOfunctions}). For simplicity we stick here to the more compact relation (valid if refraction is neglected).

\begin{equation}\label{eq:MOfunctions}
\Delta\delta(\omega) - i\Delta\beta(\omega) =  \frac{c}{\omega d_{t}}\left[{-\Phi_{F}(\omega)+i(D(\omega))}\right], 
\end{equation}
here $c$ is the  speed of light in vacuum and $ d_t $ is the total thickness of the sample at a given angle of incidence. 
See the supplementary material for further experimental details, derivation of the equations and impact of the refraction correction.
We note that the term MO-functions is used to describe what is traditionally called magneto-optical constants, in order to emphasize the frequency dependence of these complex functions which are analytic in the upper-half frequency plane.

Theoretically, the MO-functions are calculated by first performing a ground-state calculation using density functional theory (DFT) within the local spin density approximation (LSDA) for the exchange-correlation potential\cite{lda}. A single shot $GW$ calculation\cite{gw} is then employed to determine the position and width of the deep lying (low in energy) 3$p$ states. The Kohn-sham 3$p$ bands are then scissor shifted and broadened to mimic the $GW$ spectral function.  Subsequently, the response function is calculated using these Kohn-Sham states and the time-dependent extension of DFT, the so called time-dependent density functional theory (TD-DFT)\cite{RG84}. The linear response equation of TD-DFT reads\cite{my-book}:
\begin{equation} \label{chi} 
\varepsilon^{-1}(\omega)= 1+\chi_0(\omega)\left[1-(v+f_{\rm xc}(\omega))\chi_0(\omega)\right]^{-1}
\end{equation}
where $\varepsilon$ is the dielectric tensor,  $v$ is the Coulomb potential, $\chi_0$ the non-interacting response function, $f_{\rm xc}$ the exchange-correlation kernel. Electron-hole correlations, which describe the excitonic effects, can be treated by correct choice of this kernel\cite{sharma11}. The dielectric tensor is related to the experimental MO-functions by the following relation\cite{Oppeneer_hmm}:

\begin{equation} \label{beta} 
\Delta \delta(\omega) - i \Delta \beta(\omega) = 0.5i \varepsilon_{xy}(\omega)\left[\varepsilon_{xx}(\omega)\right]^{-1/2}
\end{equation}
All calculations are performed using the state-of-the art full-potential linearized augmented plane wave\cite{singh} method as implemented in the Elk code\cite{elk,param}. 

\begin{figure*}
\centering
\includegraphics[width=\textwidth	, clip]{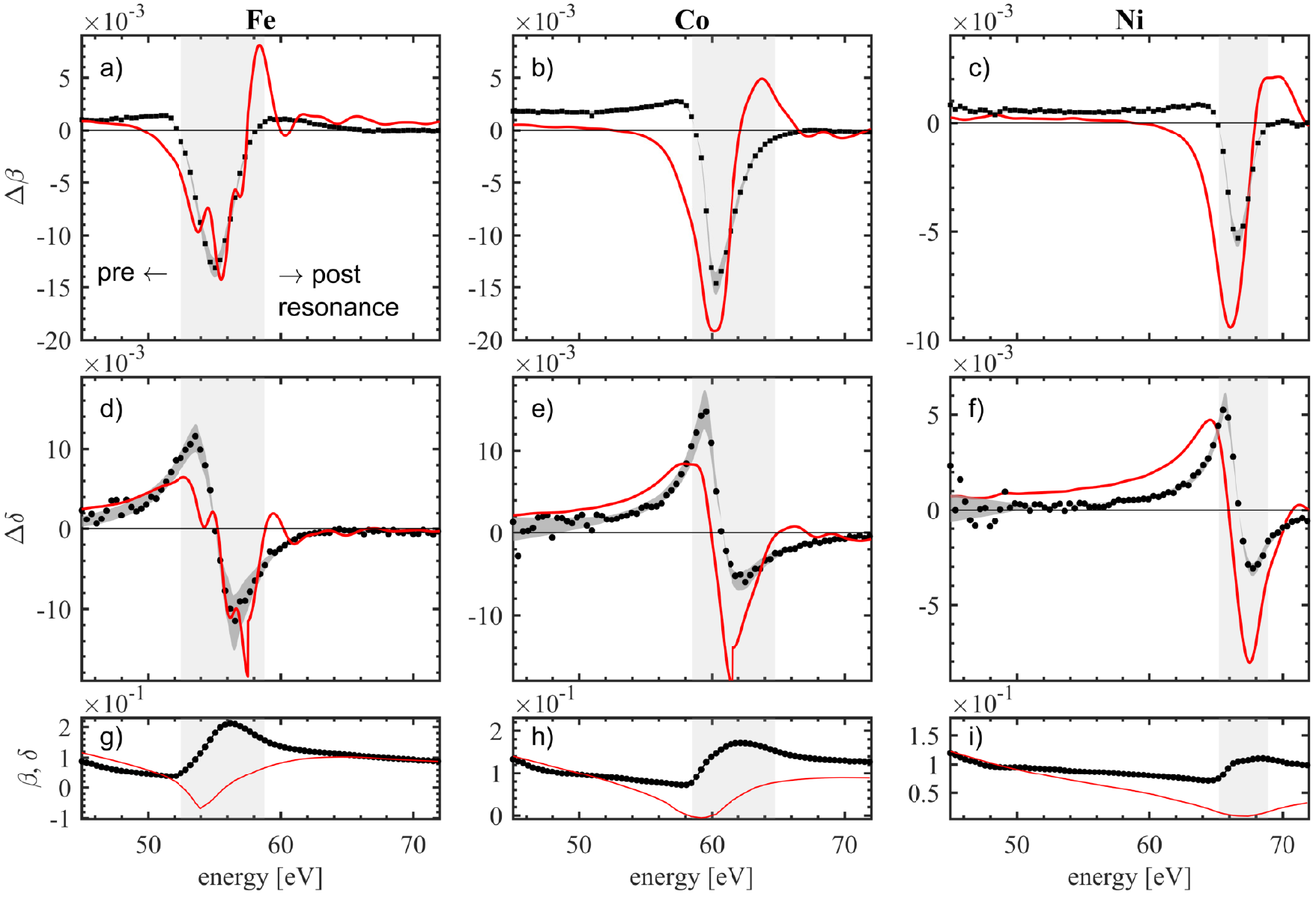}
\caption{\label{et1} Measured (black dots) and calculated (solid red line) MO-functions. The area behind the data points is the error band. The regions for resonant 3$p$-3$d$ transitions (determined by the X-ray absorption spectrum) are shaded in gray.
The absorptive part ($\Delta \beta$) is shown in panels (a,b,c) for Fe, Co, and Ni, respectively. Panels (e,f,g) display the the dispersive part ($\Delta \delta$). The optical functions that were used to correct for refraction are shown in panels (g)-(i). The theoretical data is calculated using $GW$ corrected Kohn-Sham bands and by solving TD-DFT linear response equation including local fields and excitonic effects. }
\end{figure*}
%
In Fig.~\ref{et1} we present both the experimentally determined real and imaginary
parts of the MO-functions and the \emph{ab-initio} calculated values
for Fe, Co, and Ni. The experimental data are corrected
for the grazing incidence angle $ {\Theta_{i}}$ and for refraction due to non zero $\delta$ and $\beta$ (see supplement for details).
To the best of our knowledge, this is the first measurement of \emph{both} the real and imaginary part of the MO-functions at the 3$p$ resonances for Fe, Co, and Ni, eliminating the errors associated with Kramers-Kronig analysis. 
The main error sources in the experiment were uncertainties in the
thickness of the magnetic films and  the angle of incidence. 
We estimate $ \Delta d = \pm 0.5$\ nm  and $\Delta\Theta_i \pm 2^{\circ}$.  
Especially  in the Faraday measurement the statistical error is larger: here the low reflectance (R$\approx 1-2\%$) of the Au mirror results in a lower signal to noise ratio which becomes particularly noticeable towards lower photon energies, when the beamline flux decreases rapidly. The error-bands in Fig.~\ref{et1} reflect both the systematic and statistical errors. 

From the experimental data on  the dichroic, absorptive part $\Delta\beta$ (Fig.~\ref{et1} a-c) we make two observations. 
(1) All three elements show a significant off-resonant signal ($ \approx 10\% $ of the maximum) extending from the lowest photon energy measured, i.e. 45~eV, up to the respective absorption edge. In the case of Ni, the off-resonant signal is detected up to 20~eV below the absorption edge. In Fe,  we additionally observe a sizeable off-resonant signal above the absorption edge, extending up to 65~eV.  \blk
(2) The maximum absolute amplitude of $\Delta\beta$ is smaller in Fe than in Co, in contrast to the results at the $L$-edges\cite{Chen1995}. In contrast to earlier data\cite{Valencia2006} obtained via a Kramers-Kronig inversion, we find that at the $M$-edges the amplitude of the XUV-MCD signal does \emph{not} scale with the atomic magnetic moment.


The measurement of the dispersive part of the MO-functions, $\Delta\delta$ (Fig.~\ref{et1} d-f), reveals a very symmetric bipolar curve in Fe, while in Co, and Ni the amplitude below the resonance energy is about two times of the one above. We find the maximum positive amplitude in Co. The region where $\Delta \delta$  deviates from zero extends to at most 12~eV around the zero crossing of $\Delta \delta$ at resonance.
Together with the off-resonant signal in the absorptive part, this finding is of great significance when  investigating multilayer-samples and alloys containing these elements, as their MO signals are spectrally overlapping. Thus, the exact knowledge of the MO-functions allows to predict the spectral shape and signal strength and hence elemental contributions expected in all XUV magneto-optical effects such as the magneto-optical Kerr Effect (MOKE) in transverse, longitudinal or polar geometry as well as for XUV-MCD (e.g.\cite {La-o-Vorakiat2009,Mathias2012,Turgut2013,Rudolf2012,Willems2015}). 
Moreover, in XUV scattering and imaging experiments, the choice of the photon energy not only determines the overall signal strength, but also the relative contribution of absorption vs. phase contrast\cite{Scherz2007}.

The parameter free TD-DFT spectra in Fig.~\ref{et1} are  in good agreement with the experimental data for all three elements.
Note that both theory and experiments are plotted on the same absolute scales for photon energies and MO-function magnitudes, without any adjustments. 
In order to analyze the reason behind past discrepancies and the significantly improved level of agreement between theory and experiment in the present work, and to explore the origin of various features in the XUV-MCD spectra, we probe our theoretical data in more detail. For this we discuss the three energy regions marked in Fig.~\ref{et1} (pre- , at and post-resonance) separately.

\begin{figure}[h]
	\includegraphics[ clip]{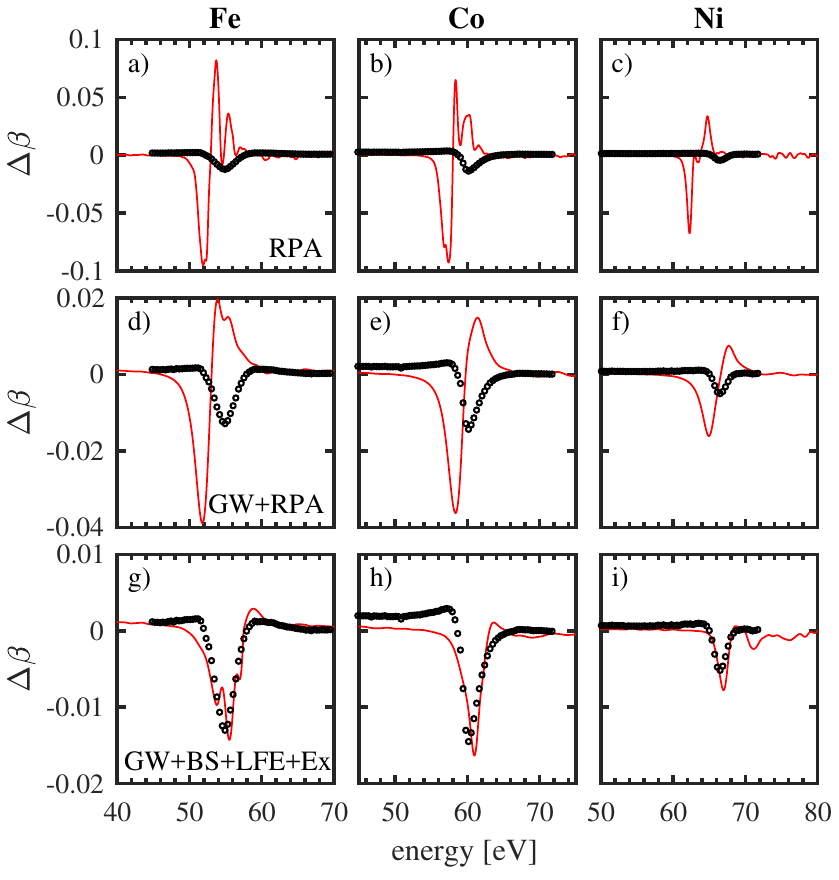}
	\caption{\label{T2} Experimental (black dots) and theoretically calculated (red solid lines) $\Delta \beta$. Results obtained using Kohn-Sham bands and RPA i.e. by using $f_{\rm xc}=0$ in Eq.~(\ref{chi}) are shown for (a) Fe, (b) Co, and (c) Ni. Results obtained using $GW$ corrected Kohn-Sham bands to account for correct position and width of 3$p$ states are presented for (d) Fe, (e) Co, and (f) Ni. Data obtained using $GW$ corrected Kohn-Sham bands, by using $f_{\rm xc}=f^{\rm boot-strap}_{\rm xc}$ to account for excitonic effects, by including local field effects by treating Eq.~(\ref{chi}) as a matrix equation and by reducing the 3$p$ exchange splitting by 60\% are presented for (g) Fe, (h) Co, and (i) Ni.} 
\end{figure}

\begin{figure}[h]
\centering
\includegraphics[ clip]{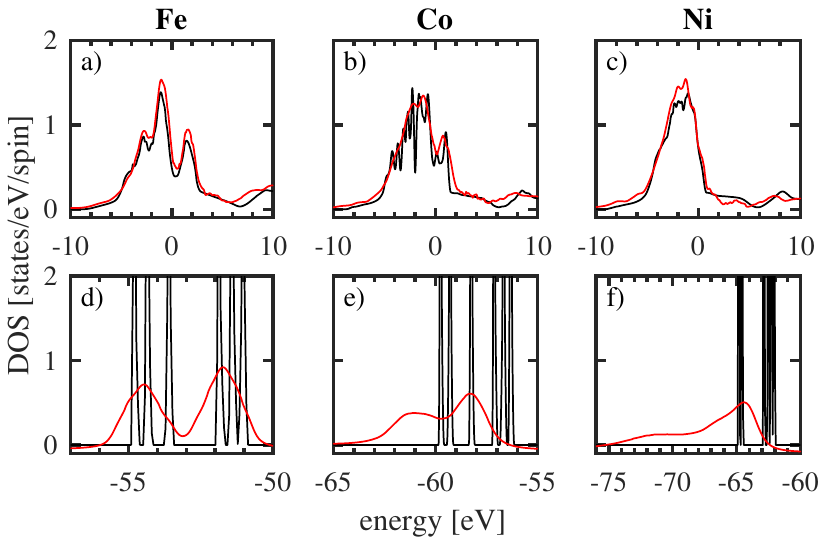}
\caption{\label{T1} Spin averaged Kohn-Sham (black lines) density of states (in states/eV/spin) and $GW$ spectral function (red lines) as a function of energy for the 3$d$ (a-c) and the 3$p$ (d-f) states of Fe, Co, and Ni.}
\end{figure}
%

At the resonance, experimental data shows a well defined peak (at 55.1~eV for Fe, 60.3~eV for Co and 66.6~eV for Ni). Calculations performed by using $f_{\rm xc}=0$ in Eq.~(\ref{chi}) (also known as the random phase approximation (RPA)) show significantly red-shifted peaks with strongly overestimated amplitudes (see Fig.~\ref{T2} (a)--(c)). 
In the past it has been speculated\cite{Turgut2016} 
that underpinning this discrepancy could be missing many-body, core-hole and excitonic effects in the Kohn-Sham band-structure which would (i) shift the 3$p$ states to lower energies and (ii) broaden these 3$p$ states. In order to investigate this we have performed fully spin polarized $GW$ calculations to determine the spectral function for bulk Fe, Co, and Ni. As expected, the many-body corrections have the effect of red-shifting the 3$p$ states as compared to the Kohn-Sham bands (see Fig.~\ref{T1} d-f). The many-body effects also lead to a finite width\cite{param2} of the 3$p$-states the values of which,  are in close agreement with the experimental work of Nyholm et al.\cite{Nyholm1981}.  In contrast, the $GW$ spectral function close to the Fermi level is almost the same as the Kohn-Sham density of states (see Fig.~\ref{T1} a-c).
With this information in hand we can now correct the Kohn-Sham bands by red shifting and broadening the 3$p$ states such that the Kohn-Sham density of states resemble the $GW$ spectral function. This results in an energy shift of the resonant peak in the response functions by 0.1~eV for Fe, 1.1~eV for Co, and 2.7~eV for Ni. While this brings the calculated resonance closer to the experiments (see Fig.~\ref{T2} (d)--(f)), the discrepancy is still  substantial for position, shape and magnitude of the peak.

In order to find the missing piece in the calculation of the response function, we turn again to Eq.~(\ref{chi}) and we note that $\chi_0$ is a matrix in reciprocal space of vectors {\bf G}; this is due to the fact that an external perturbation of the type $e^{i({\bf G}+{\bf q}) \cdot {\bf r}}$ generates a response in the density of the form $e^{i({\bf G'}+{\bf q})\cdot{\bf r}}$. Inversion of the matrix in {\bf G} space allows for inclusion of the microscopic components known as the local field effects (LFE)\cite{Aryasetiawan1992,Vast2002,my-book,sharma11,ar,bs2}. These LFE have been ignored in all the past calculations\cite{Valencia2010,Turgut2016, Zusin2018} for Fe, Co and Ni, as well as in the theoretical data shown in Fig.~\ref{T2} (a)--(f). This approximation is in addition to the setting of the exchange correlation kernel, f$_{\rm xc}$, to zero in Eq.~(\ref{chi}), which entails the neglect of electron-hole correlation (i.e. excitonic effects) in the response function.

In the following, we do include LFE by solving Eq.~(\ref{chi}) as a full matrix equation (we needed a matrix of $70 \times 70$ for convergence) and include excitonic effects by using the so called boot-strap approximation for the kernel\cite{sharma11,bs2,ar}. The results obtained using this advanced, and parameter free treatment (i.e. $GW$ corrected Kohn-Sham bands + LFE + excitons) are shown in Fig.~\ref{et1} together with the experimental data for Fe, Co, and Ni, respectively.  It is clear from these results that the TD-DFT spectra are now in excellent agreement with experiments in terms of both the position of the peak as well as its amplitude. We also note that, consistent with the experimental results, the maximum of the $\Delta \beta$ magnitude is smaller in Fe than in Co. 
This dramatic change in the quality of agreement  is, however, almost entirely due to the inclusion of LFE; calculations performed using $f_{\rm xc}=0$ but including LFE are almost indistinguishable from those using $f^{\rm boot-strap}_{\rm xc}$. These results are a clear indication that in calculations of the resonance peak in $M$-edge XUV-MCD spectra and similar observables depending on the MO-functions the LFE are of crucial importance. The $GW$ corrections are highly material dependent (almost negligible in Fe and have large effect in Ni), and excitonic effects are of no importance.
These results were to be expected since the 3$p$ and 3$d$ orbitals show a significant overlap in space. Therefore, according to the work of Aryasetiawan et al.\cite{Aryasetiawan1992} and Vast et al.\cite{Vast2002}, LFE are expected to be large. For the same reason  
LFE are small for $L$-edges, due to the small overlap of 2$p$ and 3$d$ states.  

Turning to the post-edge spectral region one notes a large peak in the positive $\Delta \beta$ direction in the calculated data which is absent in the experimental results. To investigate the origin of this peak we focus on the exchange splitting of the 3$p$ states, which is overestimated\cite{param3} by local spin density approximation\cite{blugel,karlsson,wannier}.
The spectra calculated upon reducing this splitting by 60\% (data shown in Fig.~\ref{T2}(g)-(i)) are in even better agreement with the experimental data. This is a clear indication that to treat post-edge spectra one requires not only LFE and $GW$ corrections but also improved ground-state spin-density functionals to correctly describe the exchange splitting of semi-core states. We hope that present results would stimulate future research in this direction.

As for the pre-edge part of the spectra there are two questions that remain to be addressed: what leads to the finite weight of $\Delta \beta$ in the pre-edge energy region in particular for Co and why do experiments show strikingly different behavior as compared to theory namely that there is a rise of $\Delta \beta$ just below the resonance. The answer to the first question we found to be the excitations from around the Fermi level to high lying unoccupied states. Calculation without inclusion of high lying states, more than 20~eV above the Fermi level, show that the finite weight below the resonance vanishes. Despite our best attempts we were unable to find the reason behind the peak just below the resonance.

To conclude, we provide accurate measurements and a reliable theoretical description of the magneto-optical functions for 3$d$ transition metals in the XUV regime. Finding a good agreement between the two allows us to ascribe physical processes leading to the features in the experimental data. This puts 3$d$ transition metal $M$-edge spectra - irrespective of their absorptive, dispersive or mixed character in the specific spectroscopy employed -   in the category where underlying physics can be entirely and accurately deciphered in a static and dynamic situation alike. Given the increased availability of femtosecond XUV laboratory sources, we expect this approach to be extremely useful in future studies of femtomagnetism and spintronics. 

\begin{acknowledgments}
	We thank P. Oppeneer for fruitful discussions. FW thanks R. Abrudan and K. Bauer from Helmholtz Zentrum in Berlin for technical support. 
	SS, CvKS and SE would like to thank DFG for funding through TRR227 projects A02 and A04. 
\end{acknowledgments}

\clearpage

	\section{SUPPLEMENTARY MATERIAL: Magneto-Optical Functions at the 3$p$ resonances of Fe, Co, and Ni: \emph{Ab-initio} description and experiment}
\section{Derivation of magneto-optical functions}
Peter Oppeneer developed a description of the Faraday effect in his book \cite{Oppeneer_hmm}. The derivation here follows his description and is adapted to link the measured observables to the magneto-optical (MO) functions.
The magnetic polarization of a medium influences light propagation
through the material. For circularly polarized light modes this is
described by the complex index of refraction 
\begin{equation}
n_{\pm}=1-(\delta\pm\Delta\delta)+i(\beta\pm\Delta\beta)\label{eq:complex_index_refraction}
\end{equation}
with the dispersive and absorptive part of the optical constants $\delta$
and $\beta$ and the magneto optical (MO) functions $\Delta\delta$
and $\Delta\beta$. The subscript + (-) denotes whether the helicity
of the incident light is parallel (anti-parallel) with respect to
the magnetization direction of the sample. The index of refraction
can also be referred to diagonal and off-diagonal entries of the permittivity
tensor $\epsilon$.
\begin{equation}
n_{\pm}=\sqrt{\epsilon_{xx}\pm i\epsilon_{xy}}\label{eq:n_permittivity}
\end{equation}
We calculate the ratio between two counter rotating circularly polarized
light modes to obtain the expressions for the MO functions. The circularly
polarized electric field modes $\mathbf{E}_{\pm}$transmitted through
a sample with total thickness $d_{t}$ (we distinguish between total
and effective thickness $d_{\mathrm{eff}}$. Both quantities include the angular dependence of the experiment as discussed below) can be written as 

\begin{equation}
\mathbf{E}_{\pm}=\frac{1}{\sqrt{2}}E_{0}(\mathbf{e}_{x}\pm i\mathbf{e}_{y})e^{i\frac{\omega}{c} d_{t}}_{\pm}\label{eq:field_modes}
\end{equation}
where $c$ is the speed of light, $\omega$ is the angular frequency,
$E_{0}$ is the electric field amplitude, $\mathbf{e_{x,y}}$are the
direction of the electric field and $\mathbf{z}$ is the propagation
direction. Hence the ratio of the field modes depends on the difference 
$n_{+}-n_{-}=-2(\Delta\delta-i\Delta\beta)$
and can be expressed as the product of an
amplitude and a phase factor.
\begin{eqnarray}
\frac{E_{+}}{E_{-}}&=&e^{i\frac{\omega}{c}d_{t}(n_{+}-n_{-}}\\
\underbrace{\frac{|E_{+}|}{|E_{-}|}}_{\mathrm{amplitude}}\cdot\underbrace{e^{i(\alpha_{+}-\alpha_{-})}}_{\mathrm{phase}} &=&e^{-\frac{\omega}{c}d_{t}2\Delta\beta}\cdot e^{i\frac{\omega}{c}d_{t}2\Delta\delta}
\label{eq:field quotient}
\end{eqnarray}
We apply the logarithm to Eq. \ref{eq:field quotient} and rewrite
the phase factor as the Faraday rotation angle $\Phi_{F}=\frac{1}{2}(\alpha_{+}-\alpha_{-})$
The logarithm of the
amplitude ratio can be defined as the the magnetic circular dichroism,
$D=\frac{1}{2}\ln\left(\frac{|E_{+}|}{|E_{-}|}\right)=\frac{1}{4}\ln\left(\frac{I_{+}}{I_{-}}\right)$,
with the intensities, I, of the transmitted, circularly polarized light pulses. For Eq. \ref{eq:n_permittivity}
a Taylor expansion around $\epsilon_{xy}=0$ leads in first order
to $n_{+}-n_{-}\approx i\epsilon_{xy}/\sqrt{\epsilon_{xx}}$.
With Eq. \ref{eq:field quotient} we end up with the expressions for the MO functions: 
\begin{eqnarray}
\Delta\delta-i\Delta\beta & = & \frac{c}{\omega d_{t}}(-\Phi_{F}+iD)\label{eq:MOfunctions}\\
\Delta\delta-i\Delta\beta & \approx & -\frac{i}{2}\frac{\epsilon_{xy}}{\sqrt{\epsilon_{xx}}}\label{eq:MOpermittivity}
\end{eqnarray}

Analogously to the derivation above, the product of the field modes
(Eq. \ref{eq:field_modes}) leads to the expressions for the optical
constant $\beta$. 

%

\begin{equation}
\beta  =  -\frac{c}{2\omega d_{\mathrm{eff}}}\ln\left(\frac{2\cdot|E_{+}|\cdot|E_{-}|}{E_{0}^{2}}\right)
\label{eq:beta-1}
\end{equation}
Equations (\ref{eq:MOfunctions} and \ref{eq:beta-1}) are valid for normal
incidence of the light on the sample. In the experiment the grazing
incidence angle, $\Theta_{i}$, for in plane magnetized samples was
$50^\circ$ (compare Fig.~\ref{fig:static_experimental_setup}). Therefore at oblique
incidence angles a correction for the effective thickness $d_{\mathrm{eff}}$
of the sample has to be introduced. A second correction is needed to account for the projection of the magnetization, ($\mathrm{m_{||}}$),
on the propagation direction of the light $z$. 

A geometrical consideration, see Fig.~\ref{fig:static_experimental_setup} panel (d),
leads to $m_{||}=m_{0}\cos\Theta_{t}$ and $d_{\mathrm{eff}}=d_{0}/\sin\Theta_{t}$
. Hence, the total effective thickness of the in plane magnetized
sample accounts to 
\begin{equation}
d_{t}=m_{||}\cdot d_{eff}=\frac{d_{0}}{\tan\Theta_{t}}\label{eq:oblique_angle}
\end{equation}

For spectroscopy at the M-edges where $n\neq1$, a third
correction due to refraction at the vacuum to medium interface is
necessary. We assume that $\tilde{n}=1-\delta+i\beta$ deviates slightly from 1 (the MO functions are one order of magnitude smaller and are here neglected). Consequently the angle of the refracted beam is slightly smaller
than the angle of incidence (for $\delta>0$) (grazing). We can write $\Theta_{t}=\Theta_{i}-\gamma$ and assume
a small value for $\gamma$. A Taylor expansion to first order leads
to:
\[
\tan\Theta_{t}=\tan(\Theta_{i}-\gamma)\approx\tan\Theta_{i}-\frac{\gamma}{\cos^{2}\Theta_{i,}}
\]

We use Snell's law to rewrite $\gamma\approx\frac{1-\tilde{n}}{\tilde{n}\tan\Theta_{i}}\approx\frac{\delta-i\beta}{\tan\Theta_{i}}$ (here, for the denominator in the last step, $\tilde{n}\approx1$ was assummed) and get
\begin{equation}
\tan\Theta_{t}\approx\tan\Theta_{i}\left(1-\frac{\delta-i\beta}{\sin^2\Theta_{i}}\right)\label{eq:approx_theta}
\end{equation}

Substituting $\tan\Theta_{t}$ in Eq. \ref{eq:oblique_angle} with
the expression from Eq. \ref{eq:approx_theta} and combining it with
Eq. \ref{eq:MOfunctions} leads to the expressions for the MO functions for oblique angles of incidence at the M-edges.

\begin{eqnarray}
\Delta\delta & \approx & \frac{c}{\omega d_{0}}\cdot\tan\Theta_{i}\left[-\phi_{F}+\frac{\delta\Phi_{F}-\beta D}{\sin^2\Theta_{i}}\right]\label{eq:deltadelta_corr}\\
\Delta\beta & \approx & \frac{c}{\omega d_{0}}\cdot\tan\Theta_{i}\left[D-\frac{\delta D-\beta \Phi_{F}}{\sin^2\Theta_{i}}\right]\label{eq:deltabeta_corr}
\end{eqnarray}

\section{Experimental method}
\subsection{Samples}
The samples were prepared by magnetron sputtering at Max Born Institute, Berlin.
The growing parameters were:
Base pressure 	$4.2 \cdot 10^{-7}$\ mbar,	
Argon Baratron pressure 	$2.1 \cdot 10^{-3}$\ mbar ,
growth rate: 0.16 Å/s.	
We deposited 15 nm thick layers of Fe, Co and Ni on  20 nm thick $\mathrm{Si_{3}N_{4}}$- membranes provided
by Silson Ltd. A 3 nm Al layer was added for oxidation protection. We assume that the Al layer is fully oxidized.
The error in the layer thickness is estimated to be $ \Delta d = 0.5\ nm$. This value is confirmed via a quantitative XPS-measurement.
All samples have their easy magnetization axis in the sample plane. They exhibit a rectangular hysteresis with a coercivity
of less than 10~mT.

\subsection{Experimental setup}
The experiment was performed at the undulator beamline UE112-PGM1 with variable polarization and energy resolution $\mathrm{E/\Delta E=30000}$ in
the range from 20-150 eV. The ALICE end-station \cite{Abrudan2015} was complemented with a home build polarization analyzer chamber (PAC).	
This setup allowed to measure Faraday rotation and magnetic circular dichroism (MCD) spectra in the same experiment. For this, the helicity had to be switched from linear to circular and the photo diodes were exchanged.\\
The PAC consists of an Au mirror, mounted
close to the Brewster angle at 45° and a photo diode. Mirror and detector form a unit that can be rotated around of the incoming beam by the angle $\alpha$.

\begin{figure}
	\begin{centering}
		\includegraphics[width=8.6cm]{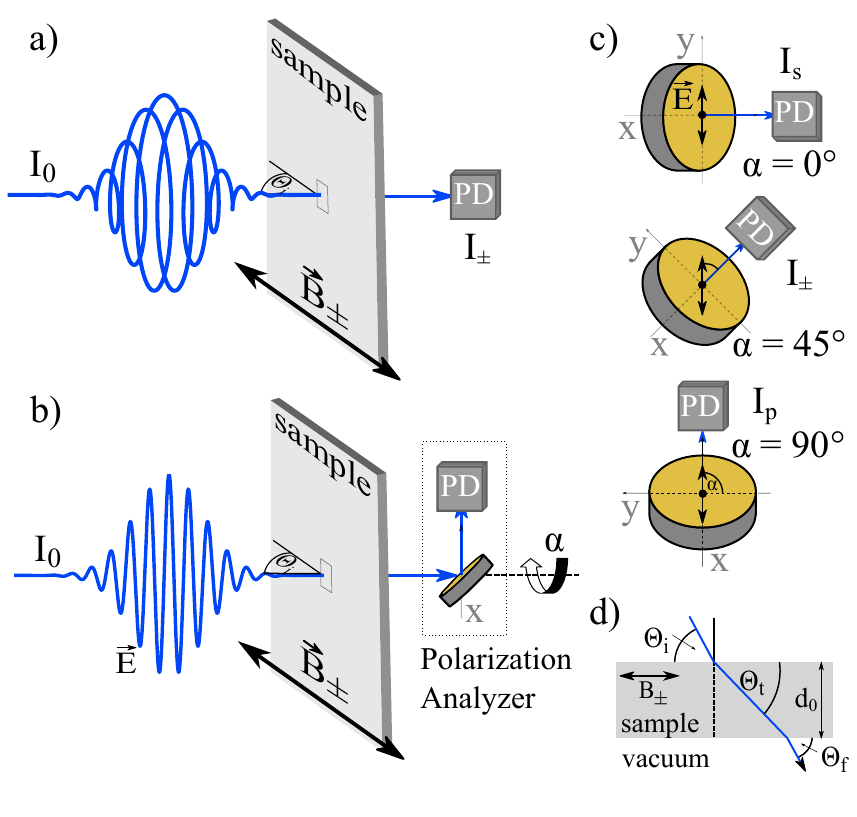}
		\par
	\end{centering}
	\caption{\label{fig:static_experimental_setup} Experimental setup for the
		measurement of $\Delta\beta$ (a) via MCD and $\Delta\delta$ in (b)
		via Faraday rotation. (a) The transmitted intensity of circularly
		polarized light is measured as a function of photon energy with a photo diode (PD) for two magnetization directions. (b) The Faraday rotation is determined with a polarization analyzer consisting of
		an Au mirror and PD. Here, linearly polarized light is transmitted through the sample and is then reflected on the Au mirror close
		to its Brewster angle at 45°. The intensity is recorded for both
		magnetization directions. The analyzer unit can be rotated by the
		angle $\alpha$ around the axis of the incident light. In
		both (a) and (b) the sample and incident light enclosed the gracing angle
		$\Theta_{i}$. In this way the in-plane magnetization has a component
		(anti-)parallel with the k-vector of the radiation. An electromagnet
		generates an external \textbf{B} field that saturates the sample in
		both field directions ($\pm$). In panel (c) three different
		analyzer positions are displayed. The angle $\alpha$ is defined as the
		angle between the polarization plane and the x axis of the analyzer
		unit. The reflected intensity from the Au is maximum for $\alpha=0\text{°}\,(I_{s})$
		and minimum for $\alpha=0\text{°}\,(I_{p})$. The photon
		energy dependent Faraday rotation spectrum is recorded at $\alpha=45{^\circ}$.
		panel (d) illustrates how refraction leads to an effective angle
		of transmission $\Theta_{t}$ in the sample.}
\end{figure}

\subsection{Setup characterization}
For the experiment it is important to know the degree of polarization of the probing light as well as the characteristics of the polarization analyzer mirror.
For the latter we measured the reflectivity for linearly polarized light for different analyzer angles at various photon energies.  

The normalized difference of maximum and minimum reflectivity for linearly polarized light gives the analyzing power of the reflection P \cite{Bahrdt2010}.
\begin{equation}
P(\omega)=\frac{I_{S}(\omega)-I_{P}(\omega)}{I_{S}(\omega)+I_{P}(\omega)}\label{eq:polarization_power}
\end{equation}
with $I_{S,P}$ being the measured intensities at $\alpha= 0^\circ,\ 90^\circ$.

The polarization power is a function of the photon energy ($\hbar\omega$).

\subsection{Characterization of the degree of polarization }

The polarization degree of the beamline UE112-PGM1 is defined by the undulator settings. There are tabulated values, for a particular polarization at the sample. These values also account for the polarization changes that are introduced when the polarized beam generated in the undulator is subsequently transported via gracing incidence on the beamline optics to the sample \cite{Bahrdt2010}.  
%
%
%
%
%
We confirmed a polarization degree of $P_{lin}=0.99 $ and $P_{circ}=0.99 $ by measuring the analyzer angle dependent reflectivity $R(\alpha) $ for both linearly and circularly polarized light modes at different photon energies.


\subsection{Data analyis and error discussion}

Figure \ref{fig:static_experimental_setup}(a) shows the experimental
setup for the MCD measurement. The intensity $\mathrm{I_{\pm}}$ of the transmitted circularly polarized beam is measured
with an XUV-sensitive photo diode for two opposite magnetization directions.
We repeated this measurement for each photon energy in the range between	45-72 eV. From this measurement we retrieved the magnetic circular dichroism D as derived above.
\begin{equation}
D=\frac{1}{4}\ln\left(\frac{I_{+}}{I_{-}}\right)
\end{equation}

Note, we defined D in a way that it shares the same pre-factors as the Faraday rotation angle $\phi_{F}$ in Eq.~\ref{eq:MOfunctions}. Hence, it is directly proportional to the material and photon energy dependent absorptive part of the MO functions $\Delta\beta(\omega)$.\\ 
In literature the magnetic circular dichroism is often defined as the normalized difference of the intensities $A =(I_+-I_-)/(I_++I_-)$. When measured in transmission geometry, this quantity is proportional to D as long as $0.8 < I_+/I_- < 1.2$. In other cases the quantities can differ significantly. Often the dichroic signal from experiments in reflection with circularly polarized light  is also called MCD~\cite{Hochst1996}). Here the MCD spectra relate to both absorptive ($\Delta\beta$) and dispersive ($\Delta\delta$) parts via the complex Fresnel equations. Hence, MCD in reflection and MCD in transmission, despite the name, measure different observables.

In the measurement for the Faraday rotation, as sketched in Fig.\ref{fig:static_experimental_setup}(b) linearly  polarized XUV
light is transmitted through the sample, reflected on the Au mirror
and its intensity measured with an XUV diode (AXUV100G by OptoDiode). 

For the energy dependent measurement we set the analyzer angle to
$\alpha=45{^\circ}$. At this angle the rotation of the polarization
plane is translated into an intensity change. The Faraday angle can
be written as \cite{Valencia2006}
\begin{equation}
\Phi_{F}=\frac{1}{2}\arcsin\left(\frac{A}{P}\right)
\end{equation}

with A being the magnetic asymmetry of the reflected intensities $I_{R\pm}$
measured for two opposite magnetic field directions and defined as
$A=(I_{R+}-I_{R-})/(I_{R+}+I_{R-})$. 

Eq. \ref{eq:deltadelta_corr} and \ref{eq:deltabeta_corr} were used 
to calculate the MO functions $\Delta\delta$ and $\Delta\beta$ shown in Fig. 1 of the main paper.
For the refraction correction we determined values for $\beta$ from the measurement according
to Eq. \ref{eq:beta-1}. Here we accounted also for 20~nm $\mathrm{Si_{3}N_{4}}$
membrane and 3~nm Al capping layer with tabulated data from CXRO
\cite{CXROHenke}. We assumed that the Al layer is completely oxidized. \\

The main error sources in the experiment were uncertainties in the
thickness of the magnetic film and in the angle of incidence. As upper
limits we estimate $\Delta d = \pm 0.5\ nm$ and $\Delta\Theta_{i}
= \pm 2^\circ$. This introduces a systematic uncertainty of 10 \%; in comparison the statistical error while recording the intensity in the MCD measurement was significantly smaller.
In the Faraday measurement the reflection on the Au mirror attenuates the beam by 2 orders of magnitude. The  intensities measured in this case approach the order of the detector noise. Especially at lower photon energies $\hbar\omega < 53~eV $ when the flux of the beamline decreases drastically, the signal to noise ratio becomes almost one. Accordingly, the error intervals shown in Fig.1 in the main paper increase towards lower photon energies.

\section{Comparison with earlier work}
\begin{figure}
	\begin{centering}
		\includegraphics[width=\textwidth]{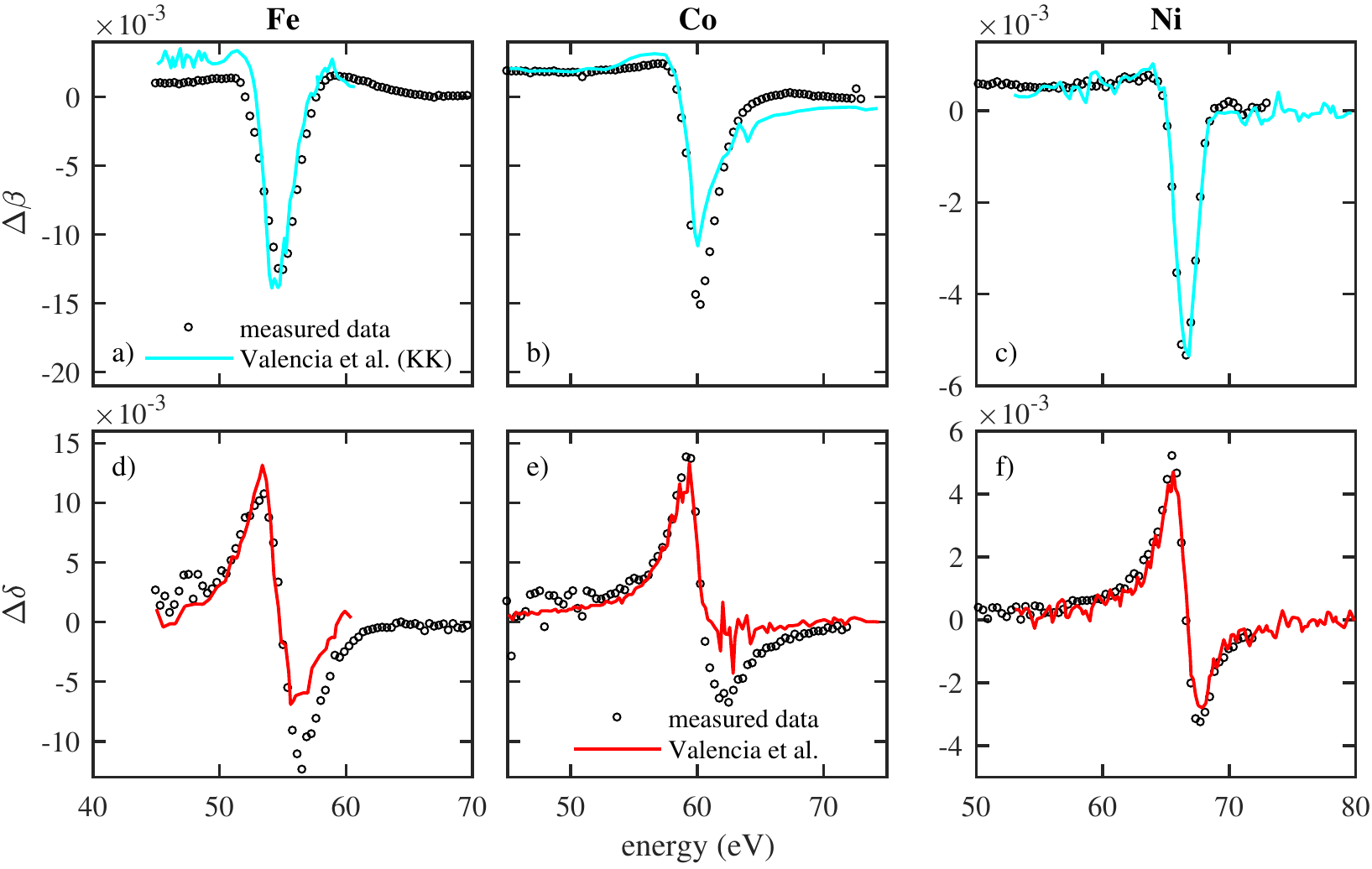}
		\par
	\end{centering}
	\caption{\label{fig:VC} Measured data (black circles) compared with the digitized data of Valencia et al. (solid curves) for Fe, Co, and Ni. Panels a)-c) show the absorptive parts (solid curves are calculated by KK-relation). In Panels d)-f) both curves are measured data.}
\end{figure}

The magneto-optical functions for Fe, retrieved from a Permalloy ($ \mathrm{Fe_{0.2}Ni_{0.8}} $) sample, Co, and Ni have been reported earlier by Valencia et al. \cite{Valencia2006}. The authors measured the dispersive and calculated the absorptive part via a Kramer-Kroning (KK) inversion. We have digitalized the data and compare it to our measurement in Fig.~\ref{fig:VC}. None of curves are refraction corrected, i.e. retrieved via Eq.~\ref{eq:MOfunctions}). Within the error bars (15-20 \% comp.~\cite{Valencia2006})  we find good agreement in the case of Ni for both dispersive and absorptive parts (panels (c) and (f)). In the case of Fe and Co we find drastic deviations in the dispersive parts (panels (d) and (e)), which appear mainly at energies higher than the resonances. In both cases this inaccuracy lead to the strong discrepancies in $\Delta\beta $.

The Fe data (red curve in panel d) was extracted from a Permalloy sample by simply drawing a line at 60\ eV and considering only the energies below this line. Outgoing from the spectral distance of Fe and Ni 3$p$ core levels of more than 10\ eV this seems to be a reasonable assumption. Though, the KK-calculated absorptive part in panel a) heavily overestimates the pre-edge part. This can be taken as proof that the supposedly far- apart lying Ni resonance overlaps, and significantly changes the a Fe signal in both, the dispersive and absorptive parts of the MO- functions.
The most striking difference between our work and Valencia et al. is that the MO functions at the M-edges do not scale with the magnetic moment, in contrast to what was reported earlier. In our work we find the largest amplitude in Co in both experimental and theoretical results.


\section{The influence of refraction correction}

\begin{figure}[h]
	\begin{centering}
		\includegraphics[width=\textwidth]{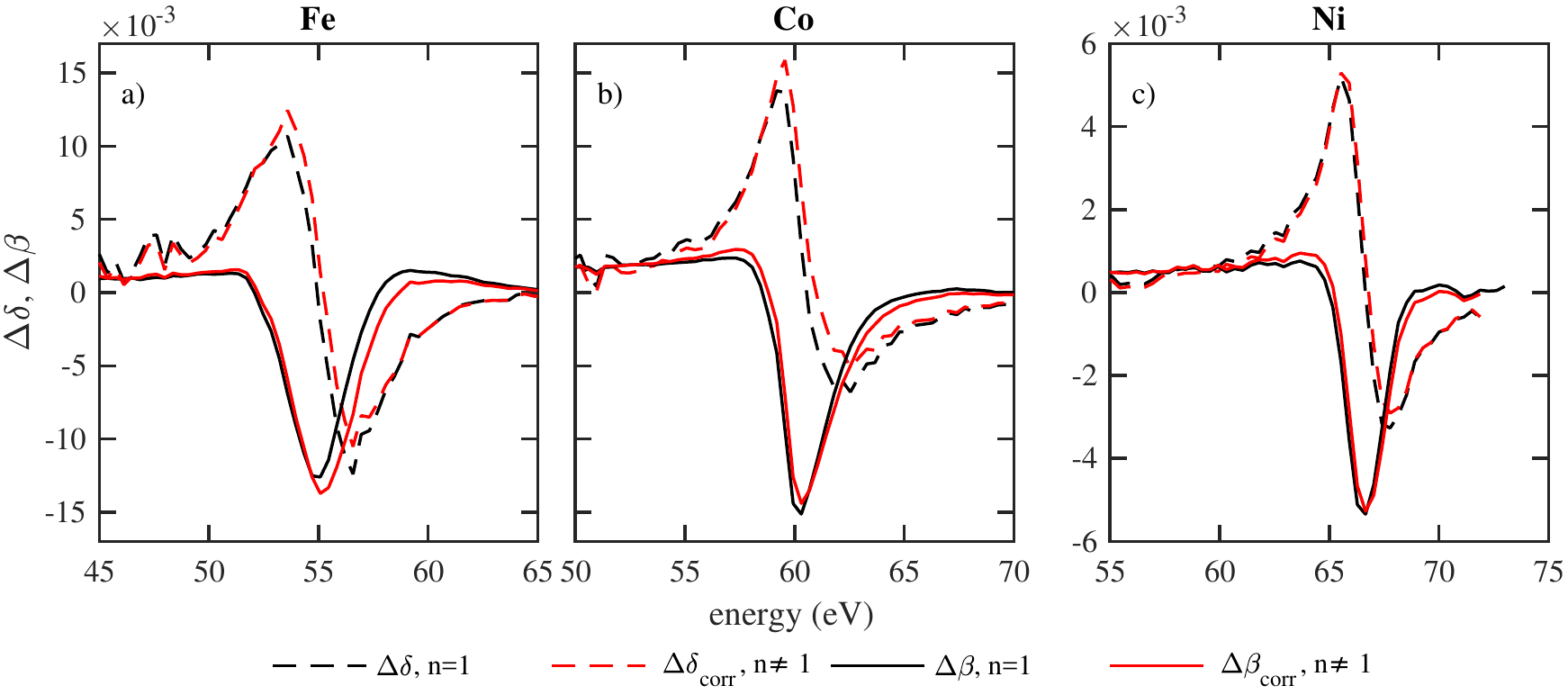}
		\par
	\end{centering}
	\caption{\label{fig:refCorr} MO-functions without (black) and with correction for refraction (red).}
\end{figure}

In Fig.~\ref{fig:refCorr} we compare the curves calculated without refraction correction (Eq.~\ref{eq:MOfunctions}) with the ones including correction obtained with Eq.~\ref{eq:deltadelta_corr} and \ref{eq:deltabeta_corr}. The corrections are significant for Fe and Co, and almost negligible in Ni.

\clearpage

\bibliographystyle{apsrev4-1}

\bibliography{MyCollection,Sang_lib}

\begin{thebibliography}{45}
\bibitem [{par({\natexlab{b}})}]{param2}%
  \BibitemOpen
  \href@noop {} {} ({\natexlab{b}}),\ \bibinfo {note} {width for $3p_{1/2}$ and
  $3p_{3/2}$ states for Fe is 1.22, 0.76~eV for Co is 1.25, 0.82~eV and for Ni
  is 1.20, 1.20~eV, respectively.}\BibitemShut {Stop}%
\bibitem [{par({\natexlab{c}})}]{param3}%
  \BibitemOpen
  \href@noop {} {} ({\natexlab{c}}),\ \bibinfo {note} {calculated exchange
  splitting for 3$p$ states is 2.7~eV for Fe, 2.2~eV for Co and 0.9~eV and for
  Ni.}\BibitemShut {Stop}%
\end{thebibliography}

\end{document}